\documentclass[aps,prd,twocolumn,superscriptaddress]{revtex4-1}

\usepackage{graphicx}

\usepackage{amsmath}
\usepackage{xfrac}
\usepackage{nicefrac}
\usepackage{subfigure}

\usepackage{xcolor}

\newcommand{\be}{\begin{equation}}
\newcommand{\ee}{\end{equation}}
\newcommand{\ba}{\begin{align}}
\newcommand{\ea}{\end{align}}

\newcommand{\x}{\vec{x}}
\newcommand{\p}{\vec{p}}
\newcommand{\pp}{\vec{p}\;\!'}

\newcommand{\q}{\vec{q}}

\newcommand{\lvac}{\langle \Omega |}
\newcommand{\rvac}{| \Omega \rangle}
\newcommand{\phia}{\phi^{\alpha}}
\newcommand{\chid}{\chi^{\dagger}}

\newcommand{\phiad}{\phia{}^{\dagger}}

\newcommand{\cG}{G}

\newcommand{\cA}{\mathcal{A}}
\newcommand{\cZ}{\mathcal{Z}}

\renewcommand{\exp}[1]{e^{#1}}
\newcommand{\expi}[2]{e^{-i #1 \cdot #2}}

\begin{document}


\title{Light Meson Form Factors at near Physical Masses}


\newcommand{\CSSM}{Special Research Centre for the Subatomic Structure
  of Matter (CSSM),\\School of Chemistry and Physics, University of
  Adelaide, South Australia 5005, Australia} 

\newcommand{\NCI}{National Computational Infrastructure
  (NCI),\\Australian National University, Australian Capital Territory
  0200, Australia} 

\newcommand{\CSIRO}{Digital Productivity Flagship, \\
	CSIRO, 15 College Road, Sandy Bay, TAS 7005, Australia}
	


\author{Benjamin J. Owen}
\affiliation{\CSSM}
\email{benjamin.owen@adelaide.edu.au}

\author{Waseem Kamleh}
\affiliation{\CSSM}

\author{Derek B. Leinweber}
\affiliation{\CSSM}

\author{M. Selim Mahbub}
\affiliation{\CSSM}
\affiliation{\CSIRO}

\author{Benjamin J. Menadue}
\affiliation{\CSSM}
\affiliation{\NCI}


\date{\today}

\begin{abstract}
The ability for most hadrons to decay via strong interactions
prevents the direct measurement of their electromagnetic properties.
However, a detailed understanding of how these resonant states feature
in scattering processes can allow one to disentangle such information
from photo production processes.  In particular, there has been
increasing interest in the determination of magnetic dipole moments
using such methods.  In a recent study \cite{Gudino:2013}, Gudi\~{n}o
{\it et al.} provide the first experimental determination of the magnetic
dipole moment of the rho meson. To facilitate a comparison with this
experimental determination, we present a calculation of the rho meson
and pion electromagnetic form factors calculated in the framework of
Lattice QCD.  Using the PACS-CS 2+1 flavour full QCD gauge field
configurations, we are able to access low $Q^2$ values at
near-physical quark masses.  Through the use of variational
techniques, we control excited state systematics in the matrix
elements of the lowest-lying states and gain access to the matrix
elements of the first excited state.  Our determination of the rho
meson $g$-factor $g_{\rho} = 2.21(8)$ is in excellent agreement with
this experimental determination, but with a significantly smaller
uncertainty.
\end{abstract}


\maketitle

\section{Introduction}

Understanding how QCD gives rise to the rich and diverse structure of
the hadrons remains an ongoing effort both theoretically and
experimentally.  Probing the electromagnetic structure remains one of
the most effective methods.  By measuring the electromagnetic form
factors of these states, we map out the distribution of charge and
magnetism within and gain valuable insight into the underlying
dynamics.

Unlike the nucleon and pion, the vast majority of QCD eigenstates are
unstable with respect to the strong interaction.  This makes
experimental determinations of their properties via conventional means
impractical due to their short lifetime.  Instead one is forced to
disentangle resonant state properties from radiative processes in
scattering experiments.  Recently, there has been interest in
extracting magnetic dipole moments
\cite{LopezCastro:2000,Kotulla:2002,Gudino:2013}.  In particular,
Gudi\~{n}o {\it et al.} \cite{Gudino:2013} have been able to provide a
determination of the magnetic dipole moment of the rho meson, the
first such experimental result for a vector meson.

It is thus timely to consider what results lattice QCD can provide as
a comparison.  The overwhelming focus within the lattice community has
understandably been on the nucleon, and to a lesser extent the pion,
where there exists a large body of experimental data
\cite{Beringer:1900}.  In comparison, there have been only a handful
of lattice studies of rho meson structure
\cite{Hedditch:2007,Gurtler:2008.PoS,Alexandrou:2002} with the
majority of these being performed in the quenched approximation to
QCD.  The only existing calculation with dynamical quarks was a
preliminary calculation \cite{Gurtler:2008.PoS} by the QCD-SF
collaboration using moderate to heavy pion masses and thus insensitive
to the chiral dynamics of full QCD.  It is also limited to a large
value of nontrivial 4-momentum transfer with $Q^2_{min} =
0.44$~GeV${}^2$.  Thus, there is an urgent need for a determination of
the rho meson form factors at low $Q^2$ and near-physical quark masses
in full QCD to allow for a direct comparison with experiment.

The progress of the past few years had lead to lattice studies near or
at physical masses and on lattice volumes significantly larger than
the states that occupy them.  Working in this new regime has reduced
the need for chiral extrapolation, but there are new systematics that
must be considered to ensure accurate results.  As quark masses
decrease, signal weakens and so one is forced to work in early
Euclidean-time regions where excited state contamination can be
significant.  Several studies have explored new approaches to properly
handle excited state contaminations with the most popular of these
being the summation method \cite{Capitani:2012,Capitani:2012.PoS} and
the variational approach
\cite{Bulava:2011,Maurer:2012,Owen:2012,Menadue:2012.PoS,Menadue:2013.PoS,Owen:2012.PoS,Owen:2013.PoS}.
Here we make use of the variational approach.  

In our recent work \cite{Owen:2012}, we demonstrated how this method
provides improved plateau quality and early onset of ground state
dominance without the need for the fine-tuning of source and sink
operators.  Furthermore, this improvement allows us to probe the
system at earlier Euclidean times giving rise to significant
improvement in the statistical uncertainties.  Preliminary studies of
electromagnetic properties \cite{Owen:2012.PoS,Owen:2013.PoS} provide
similar conclusions.

In this work we present a calculation of the rho-meson and pion form factors
utilising variational methods allowing for an accurate determination of
these quantities at near-physical masses.  Furthermore the variational
approach gives us access to the form factors of excited
states.  The use of moderately large lattice volumes allows us to
extract our results at low $Q^2$.  

For the extraction of the form factors we follow the electromagnetic
form factor formalism of Hedditch {\it et al.} \citep{Hedditch:2007}
with the necessary changes required for use within the variational
approach.

The remainder of this paper is organized as follows.  In Section 2 we
discuss the use of correlation matrix techniques in the evaluation of
hadron matrix elements.  In Section 3 we outline the framework used to
extract the pion (pseudoscalar) and rho-meson (vector) form factors on
the lattice, while Section 4 summarises the details of our lattice
simulation.  In Section 5 we present our results and finally in
Section 6 we provide concluding remarks.

\section{Theoretical Formalism}

On the lattice, the determination of hadronic observables begins with
the evaluation of the Euclidean-time two and three-point correlation
functions.  Given an interpolator $\chi(x)$ for the state in question,
the two-point function is defined as \be
\label{twopt}
\cG(\p, t) = \sum_{\x} \expi{\p}{\x} \, \lvac \, \chi(x) \, \chid(0) \, \rvac \, ,
\ee
and the three-point function
\begin{align}
\label{threept}
	\cG_{{\cal O}}(\pp ,\p, t_2, t_1) = \sum_{\x_2, \x_1} & \expi{\pp}{(\x_2-\x_1)} \, \expi{\p}{\x_1} \nonumber \\
 & \times \lvac \, \chi(x_2) \, {\cal O}(x_1) \, \chid(0) \, \rvac \, ,
\end{align}
where ${\cal O}$ is the current used to probe the state. Here we are
interested in the electromagnetic current ${\cal O} = J^{\mu}$.  In
general, this interpolator $\chi(x)$ will have overlap with all
eigenstates consistent with the given quantum numbers.  Inserting a
complete set of states between our interpolators in Eq.~\eqref{twopt}
\[
	\cG(\p, t) = \sum_{\alpha} \exp{- E_{\alpha}(\p) \, t} \lvac \, \chi(0) \, | \, \alpha, p \, \rangle \langle \, \alpha, p \, | \, \chid(0) \, \rvac \, ,
\]
we can see that the contribution to the correlation function from each
state, $\alpha$, is exponentially suppressed by its energy.  The
standard approach for examining the ground state is to work at large
Euclidean time where excited states are suppressed.  In the past this
approach has been sufficient for most quantities considered, but with
ensembles at near physical masses and studies seeking precision
determinations of hadronic observables, there has been increased
concern as to whether Euclidean time evolution is a sufficient measure
for eliminating excited state effects
\cite{Dinter:2011,Lin:2012:PoS,Alexandrou:2010:PoS}.  In this study we
select the variational approach to isolate the individual contribution
of a particular eigenstate to the two and three-point correlators.

The variational method \cite{Michael:1985,Luscher:1990} has become a
standard approach for the study of the excited state spectrum of QCD.  
The goal of this approach is to produce a set of operators $\phi^{\alpha}$ 
that couple directly to individual QCD energy eigenstates
\be
	\label{diagCond}
	\lvac \, \phia  \, | \, \beta, p \rangle \propto \delta^{\alpha \beta} \, .
\ee
The way in which this is achieved is to take an existing basis of
operators $\lbrace \, \chi_{i}(x) \, | \, i=1, \, ... \, ,n \,
\rbrace$ and construct the optimised operators $\phia$ as linear
combinations 
\be
	\label{superpos}
	\phia(x) = \sum_{i} v^{\alpha}_{i}\, \chi_{i}(x), \quad \phiad(x) = \sum_{j} \chid_{j}(x)\, u^{\alpha}_{j} \, .
\ee
Beginning with the matrix of two-point correlation functions
\[
	\cG_{ij}(\p, t) = \sum_{\x} \expi{\p}{\x} \, \lvac \, \chi_i(x) \, \chid_j(0) \, \rvac \, ,
\]
and making use of Eqs.~\eqref{diagCond} and \eqref{superpos}, both
$v^{\alpha}_{i} \, \cG_{ij}(\p, t)$ and $\cG_{ij}(\p, t) \,
u^{\alpha}_{j}$ have a recurrence relation governed by the energy of
state $\alpha$.  It follows that the necessary vectors
$v^{\alpha}_{i}$ and $u^{\alpha}_{j}$ are the solutions of the
generalised eigenvalue equations
\begin{subequations}
\begin{align}
\label{LeftEV}
	v^{\alpha}_{i} \, \cG_{ij}(\p, t_0 + \delta t) \hspace{15pt} =
        & \, \exp{-E_{\alpha}(\p) \, \delta t} \, v^{\alpha}_{i} \,
        \cG_{ij}(\p, t_0) \, , \\
\label{RightEV}
	\cG_{ij}(\p, t_0 + \delta t) \, u^{\alpha}_{j} = & \,
        \exp{-E_{\alpha}(\p) \, \delta t} \hspace{16pt} \cG_{ij}(\p,
        t_0) \, u^{\alpha}_{j} \, . 
\end{align}
\end{subequations}
It is important to note that Eqs.~\eqref{LeftEV} and \eqref{RightEV}
are evaluated for a given 3-momentum $\p$ and so the corresponding
operators satisfy Eq.~\eqref{diagCond} \textit{for this momentum only}.  The
correlator for the state $| \, \alpha, p \, \rangle$ can be obtained
from the correlation matrix by projecting with the relevant
eigenvectors
\[
	\cG(\p, t; \alpha)= v^{\alpha}_{i}(\p) \, \cG_{ij}(\p, t) \, u^{\alpha}_{j}(\p) \, .
\]
The extension to three-point correlators follows simply.  Beginning
with the matrix of three-point correlation functions
\begin{align*}
	(\cG_{{\cal O}})_{ij}(\pp, \p, t_2, t_1) = \sum_{\x_2,\x_1} &\expi{\pp}{(\x_2-\x_1)} \expi{\p}{\x_1} \\
	& \times \lvac \, \chi_i(x_2) \, {{\cal O}}(x_1) \, \chid_j(0) \, \rvac \,
\end{align*}
and having evaluated the eigenvectors with the two-point correlation
matrix, it is a simple matter of projecting the necessary eigenvectors
onto the matrix of three-point functions, where care is taken to to
ensure that the projection is done with the correct momenta for source
and sink, 
\[
	\cG_{{\cal O}}(\pp, \p, t_2, t_1; \alpha) \equiv v^{\alpha}_{i}(\pp) \, (\cG_{{\cal O}})_{ij}(\pp, \p, t_2, t_1) \, u^{\alpha}_{j}(\p) \, .
\]
This applies equally well to hadron transitions ($L \overset{{\cal
    O}}{\rightarrow} R$) where one simply projects with relevant
eigenvectors for the differing left ($L$) and right ($R$)
eigenstates. 

The application of this approach to states with explicit
spin-degrees-of-freedom is straight forward.  For a state with spin-1,
relevant to this work, the matrix of two-point correlation functions
is
\[
	\cG_{ij,\sigma \tau}(\p, t) = \sum_{\x} \expi{\p}{\x} \lvac \, \chi_{i,\sigma}(x) \, \chid_{j,\tau}(0) \, \rvac \, ,
\]
where the Roman indices denote the operator in the variational basis
and the Greek indices denote the Lorentz indices.  Following the
arguments set out above, we can construct eigenvalue equations for
each Lorentz component and define the corresponding optimised
operators as
\be
	\phia_{\sigma}(x) = \sum_{i} v^{\alpha}_{i,\sigma}\,
        \chi_{i,\sigma}(x), \:\: \phia_{\tau}{}^{\dagger}(x) =
        \sum_{j} \chid_{j, \tau}(x)\, u^{\alpha}_{j, \tau} \, ,
\ee
{\it i.e.}~we determine the eigenvectors for each Lorentz component
separately such that the operator maximally isolates the spectrum
observed in each Lorentz component
\footnote{We note that an analysis accounting for finite-volume
  systematic errors will need to first construct the spin eigenstates
  due to finite-volume induced non-degeneracies in the spectrum
  \cite{OwenYoungEtal}. }.
%
%
Thus our projected two and three-point correlation functions are
\begin{widetext}
\begin{align*}
	\hspace*{-8pt} \cG_{\sigma \tau}(\p, t; \alpha) &\equiv v^{\alpha}_{i,\sigma}(\p)\; \cG_{ij,\sigma \tau}(\p, t)\; u^{\alpha}_{j, \tau}(\p) \\
	\hspace*{-8pt} (\cG_{\cal O})_{\sigma \tau}(\pp, \p, t_2, t_1; \alpha) &\equiv v^{\alpha}_{i,\sigma}(\pp)\; (\cG_{\cal O})_{ij,\sigma \tau}(\pp, \p, t_2, t_1)\; u^{\alpha}_{j, \tau}(\p) \, .
\end{align*}

Having obtained the projected two and three-point correlation
functions, determination of matrix elements follows in the standard
way with the construction of a suitable ratio to isolate the desired
terms.  We use the ratio defined by Hedditch, {\it et
al.} \cite{Hedditch:2007}
\be
\label{Ratio}
\hspace*{-10pt}R_{\sigma}{}^{\mu}{}_{\tau}(\pp, \p; \alpha) = \sqrt{
  \frac{ \langle \cG_{\sigma \tau}^{\mu}(\pp ,\p, t_2, t_1; \alpha)
    \rangle \, \langle \cG_{\tau \sigma}^{\mu}(\p, \pp, t_2, t_1; \alpha)
    \rangle }{ \langle \cG_{\sigma \sigma}(\pp, t_2; \alpha) \rangle
    \, \langle \cG_{\tau \tau}(\p, t_2; \alpha) \rangle } } \, ,
\ee
\end{widetext}
where repeated indices are not summed over.  We note that the use of
the square root requires the sign of the result to be recovered from
the individual three-point functions.  Moreover the
covariant/contravariant placement of indices is for clarity only.
The advantage of this construction is that we have exact cancellation
of the momentum-dependent $\cZ^{\alpha}(\p)$ factors.  In the case of
the pseudoscalar pion interpolating field, the $\sigma\tau$ indices
are not required and can be ignored.

As in Ref.~\citep{Hedditch:2007,Boinepalli:2006,Boinepalli:2009} we consider both $\lbrace U
\rbrace$ and $\lbrace U^* \rbrace$ configurations in creating an
improved unbiased estimator \cite{Draper:1988}.  This requires the
evaluation of both $+\q$ and $-\q$ SST-propagators, thus incorporating
parity symmetry.  A consequence of this is that our correlators are
perfectly real or imaginary, depending on the matrix element under
consideration.  As established in Ref.~\cite{Draper:1988}, this
approach provides improved plateaus and a reduction of statistical
uncertainties beyond a naive doubling of the number of
configurations considered.

\section{Meson Form Factors}

\subsection{Pseudoscalar Mesons}

Having defined suitable operators for creating and isolating a
particular state $\alpha$, it now stands to understand how the form
factors are extracted from the ratio defined in Eq.~\eqref{Ratio}.  We
begin this discussion with $\pi$ meson.

As pseudoscalar mesons are spinless particles, the operator overlap
can be decomposed simply
\[
	\lvac \, \phi^{\alpha,\p}(0) \, | \, \pi_{\beta}(\p) \, \rangle = \frac{ \delta^{\alpha \beta}}{\sqrt{2E_{\alpha}(\p)}} \, \cZ^{\alpha}(\p) \, ,
\]
where $\cZ^{\alpha}(\p)$ enumerates the coupling strength of the
operator $\phia$ to the state $| \, \pi_{\beta}(\p) \, \rangle$.  It
follows that the two-point function can be expressed as
\be
\label{pi2pt}
	\cG(\p, t_2; \alpha) = \frac{\exp{- E_{\alpha}(\p) \, t_2}}{2E_{\alpha}(\p)} \, \cZ^{\alpha}(\pp) \, \cZ^{\alpha}{}^{\dagger}(\p)
\ee
and the three-point correlation function
\begin{align}
\label{pi3pt}
	\cG^{\mu}(\pp, \p, t_2, t_1; \alpha) = \frac{\exp{- E_{\alpha}(\pp) \, (t_2 - t_1)} \exp{- E_{\alpha}(\p) \, t_1}}{2\sqrt{E_{\alpha}(\pp) E_{\alpha}(\p)}} & \nonumber \\
 \times \cZ^{\alpha}(\pp) \cZ^{\alpha}{}^{\dagger}(\p) \, \langle \, \pi_{\alpha}(\pp) \, | \, J^{\mu}(0)& \, | \, \pi_{\alpha}(\p) \, \rangle \, ,
\end{align}
with the matrix element $\langle \pi_{\alpha}(\pp) | J^{\mu}(0) |
\pi_{\alpha}(\p) \rangle$ encoding the interaction vertex of the pion
with the electromagnetic current.  For a pseudoscalar meson, this
vertex is parametrized by a single form factor $G_{C}(Q^2)$ 
\begin{align}
\label{pivertex}
	\hspace*{-8pt} \langle \, \pi_{\alpha}(\pp) \, | \, J^{\mu}(&0) \, | \, \pi_{\alpha}(\p) \, \rangle = \nonumber \\ 
	&\frac{1}{2\sqrt{E_{\alpha}(\pp) E_{\alpha}(\p)}} \left[ p'^{\mu} + p^{\mu} \right] G^{\alpha}_{C}(Q^2) \, ,
\end{align}
where $p$ and $p'$ are the incoming and outgoing 4-momenta and $Q^2 =
-q^2$ is the space-like momentum transfer.  
Substituting Eqs.~\eqref{pi2pt}, \eqref{pi3pt} and \eqref{pivertex}
into Eq.\eqref{Ratio}, we obtain
\[
R^{\mu}(\pp, \p; \alpha) = \frac{1}{2\sqrt{E_{\alpha}(\pp)E_{\alpha}(\p)}} \left[ p^{\mu} + p'^{\mu} \right] G^{\alpha}_C(Q^2) \, .
\]
Together with our choice of kinematics, $p' = (E_{\alpha}, p_x, 0, 0)$ and $p = (m_{\alpha}, 0, 0, 0)$ with $E_{\alpha} = \sqrt{m_{\alpha}^2 + p_x^2}$, we access $G^{\alpha}_C(Q^2)$ through
\be
G^{\alpha}_C(Q^2) = \frac{2\sqrt{m_{\alpha}E_{\alpha}}}{E_{\alpha}+m_{\alpha}}\, R^{0}(\p_x, 0; \alpha) \, .
\ee

\subsection{Vector Mesons}

For a vector meson, the overlap term now includes the spin polarization
vector $\epsilon(p,s)$
\[
	\lvac \, \phi_{\sigma}^{\alpha,\p}(0) \, | \, \rho_{\beta}(\p, s) \, \rangle = \frac{ \delta^{\alpha \beta}}{\sqrt{2E_{\alpha}(\p)}} \, \cZ^{\alpha}(\p) \, \epsilon^{\alpha}_{\sigma}(p,s) \, ,
\]
which satisfies the spin sum relation
\be
\label{spinsum}
	\sum_s \epsilon^{\alpha}_{\sigma}(p,s) \epsilon^{\alpha}_{\tau}{}^*(p,s) = - \left( g_{\sigma \tau} - \frac{p_{\sigma} p_{\tau}}{m_{\alpha}^2} \right) \, .
\ee
Using these expressions, the two-point function takes the form
\begin{widetext}
\be
\label{rho2pt}
	\cG_{\sigma \tau}(\p, t_2; \alpha) = - \frac{\exp{- E_{\alpha}(\p) \, t_2}}{2E_{\alpha}(\p)} \, \cZ^{\alpha}(\p) \, \cZ^{\alpha}{}^{\dagger}(\p) \, \left( g_{\sigma \tau} - \frac{p_{\sigma} p_{\tau}}{m_{\alpha}^2} \right) \, ,
\ee
while the three-point correlation function is
\be
\label{rho3pt}
	\cG^{\mu}_{\sigma \tau}(\pp, \p, t_2, t_1; \alpha) =
        \frac{\exp{- E_{\alpha}(\pp) \, (t_2 - t_1)} \exp{-
            E_{\alpha}(\p) \, t_1}}{2\sqrt{E_{\alpha}(\pp)\,
            E_{\alpha}(\p)}} \, \cZ^{\alpha}(\p)\,
        \cZ^{\alpha}{}^{\dagger}(\p)\, 
        \epsilon'^{\alpha}_{\sigma}(p',s') \; \langle \,
        \rho_{\alpha}(\pp, s') \, | \, J^{\mu}(0) \, | \,
        \rho_{\alpha}(\p, s) \, \rangle \;
        \epsilon^{\alpha}_{\tau}{}^*(p,s) \, , 
\ee
where again the matrix element $\langle \rho_{\alpha}(\pp, s') |
J^{\mu}(0) | \rho_{\alpha}(\p, s) \rangle$ contains all of the
information governing the interaction with the electromagnetic
current.  For a spin-1 system, the interaction vertex can be described
by three independent vertex functions
\cite{Kim:1973,Arnold:1980,Brodsky:1992}
\be
\langle \, \rho_{\alpha}(\pp, s') \, | \, J^{\mu}(0) \, | \,
\rho_{\alpha}(\p, s) \, \rangle = \frac{1}{2\sqrt{E_{\alpha}(\p)\,
    E_{\alpha}(\pp)}} \, \epsilon'^{\alpha}_{\delta}{}^{*}(p',s') \;
\Gamma^{\delta \mu \gamma}(p', p) \; \epsilon^{\alpha}_{\gamma}(p,s) 
\ee
where
\[
	\Gamma^{\delta \mu \gamma}(p', p) = - \, \left \{ \, g^{\delta
          \gamma} \, [ p^{\mu} + p'^{\mu} ] \, G_1(Q^2) \, + \, [
          g^{\gamma \mu} q^{\delta} - g^{\delta \mu} q^{\gamma} ] \,
        G_2(Q^2) \, - \, q^{\delta} q^{\gamma} \frac{ p^{\mu} +
          p'^{\mu} }{2 m_{\alpha}^2} \, G_3(Q^2) \right \} \, .
\]
Taking the appropriate linear combination of these vertex functions, we arrive at the Sachs decomposition \cite{Arnold:1980,Brodsky:1992}
\begin{align*}
	G_Q(Q^2) &= G_1(Q^2) - G_2(Q^2) + (1 + \eta) \, G_3(Q^2) \, , \\
	G_M(Q^2) &= G_2(Q^2) \, , \\
	G_C(Q^2) &= G_1(Q^2) + \frac{2}{3} \, \eta \, G_{Q}(Q^2) \, ,	
\end{align*}
where $\eta = \frac{Q^2}{4m_{\alpha}^2}$.

Substituting the vertex into Eq.~\eqref{rho3pt} and applying Eq.~\eqref{spinsum}, the three-point function becomes
\[
	\cG^{\mu}_{\sigma \tau}(\pp, \p, t_2, t_1; \alpha) = \frac{\exp{- E_{\alpha}(\pp) (t_2 - t_1)} \, \exp{- E_{\alpha}(\p) t_1}}{2\sqrt{E_{\alpha}(\pp) E_{\alpha}(\p)}} \cZ^{\alpha}(\p) \, \cZ^{\alpha}{}^{\dagger}(\p) \, \cA_{\sigma}{}^{\mu}{}_{\tau}(\pp, \p) \, ,
\]
where we have grouped all covariant indices into the term
\[
\cA_{\sigma}{}^{\mu}{}_{\tau}(\pp, \p) = \left( g_{\sigma \delta} - \frac{p'_{\sigma} p'_{\delta}}{m_{\alpha}^2} \right) \Gamma^{\delta \mu \gamma}(p', p) \left( g_{\gamma \tau} - \frac{p_{\gamma} p_{\tau}}{m_{\alpha}^2} \right) \, .
\]
Using the symmetry property $\cA_{\sigma}{}^{\mu}{}_{\tau}(\pp, \p) =
\cA_{\tau}{}^{\mu}{}_{\sigma}(\p, \pp)$ and our choice of kinematics,
the ratio, defined by Eq.~\eqref{Ratio}, becomes 
\[
	R_{\sigma}{}^{\mu}{}_{\tau}(\p_x, 0) = \frac{1}{2\sqrt{m_{\alpha}E_{\alpha}}} \left( \frac{p'_{\sigma} p'_{\sigma}}{m^2_{\alpha}} - g_{\sigma \sigma} \right)^{-\sfrac{1}{2}} \cA_{\sigma}{}^{\mu}{}_{\tau}(\p_x, 0).
\]
In particular we consider
\begin{align*}
\cA_{1}{}^{0}{}_{1}(\p_x, 0) &= \frac{E_{\alpha}}{m_{\alpha}}
\left((E_{\alpha}+m_{\alpha}) \, G_C(Q^2) + \frac{2}{3}
\frac{p_x^2}{m_{\alpha}} \, G_Q(Q^2)\right) \, , \\
\cA_{2}{}^{0}{}_{2}(\p_x, 0) = \cA_{3}{}^{0}{}_{3}(\p_x, 0) &=
 \left((E_{\alpha}+m_{\alpha}) \, G_C(Q^2) - \frac{1}{3}
 \frac{p_x^2}{m_{\alpha}} \, G_Q(Q^2)\right) \, , \\
\cA_{1}{}^{3}{}_{3}(\p_x, 0) &= - \frac{E_{\alpha}}{m_{\alpha}} p_x \,
G_M(Q^2) \, , \\
\cA_{3}{}^{3}{}_{1}(\p_x, 0) &= +p_x \, G_M(Q^2) \, .
\end{align*}
As such, the form factors for the rho meson are isolated from the
following combination of ratio terms
\begin{align}
G_C(Q^2) &= \frac{2}{3} \frac{\sqrt{E_{\alpha} m_{\alpha}}}{E_{\alpha}
  + m_{\alpha}} \left( R_{1}{}^{0}{}_{1} + R_{2}{}^{0}{}_{2} +
R_{3}{}^{0}{}_{3} \right) \, , \\
G_M(Q^2) &= \frac{\sqrt{E_{\alpha} m_{\alpha}}}{p_x} \left(
R_{3}{}^{3}{}_{1} - R_{1}{}^{3}{}_{3}\right) \, , \\
G_Q(Q^2) &= \frac{m_{\alpha} \sqrt{E_{\alpha} m_{\alpha}}}{p_x^2}
\left( 2R_{1}{}^{0}{}_{1} - R_{2}{}^{0}{}_{2} -
R_{3}{}^{0}{}_{3}\right) \, .
\end{align}
\end{widetext}

\subsection{Extracting Static Quantities}

The Sachs form factors $G_C$, $G_M$ and $G_Q$ describe the
distribution of charge, magnetism and charge asymmetry within the
hadron.  In particular, the value of these functions at $Q^2=0$ define
the hadron's total charge, $q$, magnetic moment, $\mu$, and quadrupole
moment, $Q$,
\begin{subequations}
	\begin{align}
		q = \: \: e \, \: \: \: &G_C(0)\, , \\
		\mu = \frac{e}{2 m} \: &G_M(0)\, , \\
		Q = \frac{e}{m^2} \: &G_Q(0)\, ,
	\end{align}
\end{subequations}
where $m$ is the mass of the hadron.
In this calculation, we work with fixed-current SST-propagators.  This
allows us to explore the form factors of many different hadrons
without the need for additional inversions, but limits us to a single
3-momentum transfer and thus a single $Q^2$ for each quark mass.

We make use of a monopole ansatz for the $Q^2$ dependence
\be
	G_{i}(Q^2) = \left( \frac{\Lambda^2}{\Lambda^2 + Q^2} \right) G_{i}(Q^2=0) \, ,
\ee
as suggested by a Vector Dominance Model hypothesis.  Here we use a
conserved current, $G_C(Q^2=0) = 1$, and so we extract the monopole
squared-mass $\Lambda^2$ via
\be
\Lambda^2 = \frac{Q^2}{1 - G_C(Q^2)} \, G_C(Q^2) \, .
\ee
Motivated by the observed scaling behaviour for $G_E$ and $G_M$ of
the nucleon at low $Q^2$, we shall assume the meson sector displays
similar scaling for each quark sector and use this to extract a value
for $G_M(Q^2=0)$
\be
	G_i(Q^2=0) = \frac{G_i(Q^2)}{G_C(Q^2)} \, ,
\ee
to facilitate a comparison with the experimental prediction of
\cite{Gudino:2013} and model expectations.  Drawing on the monopole
ansatz, we shall also use this for $G_Q$.
 
For the mean squared charge radius, we use the standard definition from the small $Q^2$ expansion of the Fourier transform of the charge distribution which gives
\be
	\langle r^2 \rangle = -6 \frac{d}{dQ^2} \left. G_C(Q^2) \right|_{Q^2=0} \, .
\ee
Using our monopole form we have
\be
	\langle r^2 \rangle = \frac{6}{Q^2} \left( \frac{1}{G_C(Q^2)} - 1 \right) \, .
\ee

\section{Simulation Details}

Two and three-point correlation functions are evaluated following the
prescription outlined in
\cite{Boinepalli:2006,Leinweber:1990,Leinweber:1992,Boinepalli:2009}.
For this calculation we make use of the PACS-CS (2+1)-flavour
dynamical-QCD gauge field configurations \cite{Aoki:2008} made
available through the ILDG \cite{Beckett:2009}.  These configurations
are generated using a non-perturbatively $\mathcal{O}(a)$-improved
Wilson fermion action and Iwasaki gauge action on a $32^3 \times 64$
lattice with periodic spatial boundary conditions.  The scale is
determined via the static quark potential.  The value $\beta = 1.90$
gives a lattice spacing $a = 0.0907(13)$~fm resulting in a physical
volume of spacial extent $L = 2.9$~fm. We have access to 5 light-quark
masses, with the strange quark mass held fixed.  The resulting pion
masses range from 702~MeV to 156~MeV.

A fixed boundary condition is applied to the temporal direction and
our quark sources are inserted at $t_{src}=16$.  We have verified that
reflections associated with the fixed boundary are negligible at
Euclidean times greater than 16 time slices away from the boundary.
Due to the limited number of configurations at the lightest mass we
make use of multiple quark sources on each configuration.  This is
done by using two maximally separated spacial sources with a relative
temporal boundary shift of 8 time slices.  The temporal boundary is
then shifted by multiples of 16 time slices for each of these spacial
sources.  Multiple sources are also used for the next two lightest quark masses
with a single spacial position, separated temporally by 32 time
slices.  Table~\ref{tab:ensParam} provides a summary of the simulation
details for each $\kappa$-value and the corresponding value for
$m_{\pi}^2$.

\begin{table}[t]
	\centering
	\caption{Ensemble parameters used in this calculation.}
	\begin{ruledtabular}
	\begin{tabular}{cccc}
		$\kappa_{ud}$ & $m_{\pi}$ (MeV) & $N_{\rm cfgs}$ & $N_{\rm srcs}$ \\
		\hline
		0.13700 & 702 & 350 & 350  \\
		0.13727 & 570 & 350 & 350  \\
		0.13754 & 411 & 350 & 700  \\
		0.13700 & 296 & 350 & 700  \\
		0.13781 & 156 & 197 & $\sim$1600 \\
	\end{tabular}
	\end{ruledtabular}
\label{tab:ensParam}
\end{table}

The three-point correlation functions are evaluated using a sequential
source technique with the current held fixed as outlined in
\cite{Boinepalli:2006,Leinweber:1990,Leinweber:1992,Boinepalli:2009}. For
the current we use an $\mathcal{O}(a)$-improved conserved vector
current obtained using the Noether procedure with the improvement term
constructed in the form of a total four-divergence
\cite{Boinepalli:2006,Martinelli:1990}.  The current is inserted at
$t_c = 21$ relative to the quark insertion time at $t_{src} = 16$.
The resulting source-current separation would in general be too small
and so give rise to excited state contamination, but as was
highlighted in \cite{Owen:2012} our use of the variational approach
gives rise to rapid ground state dominance after the source allowing
for considerably early current insertion and smaller
source-current-sink separations.

Given our choice of correlation-function ratio, we require
SST-propagators for both $+\q$ and $-\q$ where we choose $\q =
\frac{2\pi}{L} \hat{x}$.  The current is polarized with $\mu = 3, 4$
as required by our evaluations of the form factors.  As outlined in
Ref.~\cite{Draper:1988}, the invariance of the lattice action under $U
\rightarrow U^*$ implies the link variables $\lbrace U \rbrace$ and
$\lbrace U^* \rbrace$ are of equal weight and so we choose to account
for both sets of configurations in the evaluation of our correlation
functions.  Rather than performing further matrix inversions on the
$\lbrace U^* \rbrace$ configurations, we make use of the fermion
matrix property
\[
M( \lbrace U^* \rbrace ) = \left( \tilde{C}\, M( \lbrace U \rbrace )\, \tilde{C}^{-1} \right)^*
\]
with $\tilde{C} = C\, \gamma_5$
to give us the propagators for the $\lbrace U^* \rbrace$ from the existing $\lbrace U \rbrace$ propagators \cite{Boinepalli:2006}
\[
S(x, 0; {U^*}) = (\tilde{C}\, S(x, 0; {U})\, \tilde{C}^{-1} )^* \, .
\]
For the fixed-current SST-propagators this identity takes the form
\[
S(x, 0; t, \q, \mu; {U^*}) = (\tilde{C}\, S(x, 0; t, -\q, \mu; {U})\, \tilde{C}^{-1} )^* \, .
\]
Thus, both $\q$ and $-\q$ momentum insertions are required as
discussed at the end of Section 2.  Our error analysis is performed
using a second-order jackknife where the $\chi^2 / \mathrm{dof}$ for
our fits is obtained through the covariance matrix.

As the primary goal of this work is to cleanly extract matrix elements
of the ground state, we choose to work with a small variational basis.
Our operators are local meson operators of varying widths.  This is
achieved by applying increasing levels of gauge invariant Gaussian
smearing \cite{Gusken:1989qx} to our quark sources and sinks.
The smearing procedure is: 
\begin{align}
\psi_{i}(x,t) &=\sum_{x'}\, F(x,x')\, \psi_{i-1}(x',t) \, ,
\end{align}
where
\begin{align}
F(x,x') &= {(1-\alpha)}\, \delta_{x,x'} \\
        &+ \frac{\alpha}{6}
\sum_{\mu=1}^{3} \left [ U_{\mu}(x)\, \delta_{x',x+\hat\mu} 
                + U_{\mu}^{\dagger}(x-\hat\mu)\,
                \delta_{x',x-\hat\mu} \right ],
\nonumber
\end{align}
where the parameter $\alpha=0.7$ is used in our calculation.  We use
the four different levels of smearing examined in
Ref.~\cite{Mahbub:2010} to provide an optimal basis for these
ensembles.  Table~\ref{tab:smearingRadii} lists the corresponding RMS
radii for the quark sources.
We focus on a single spin-flavour construction for the meson
interpolators and draw on the various source/sink smearing widths to
enable the efficient and accurate isolation of states in the
variational approach.
The use of a variety of widths enables the resulting operators to form 
nodal structures in the radial wave functions of
the excited states \cite{Roberts:2013} and is central to our ability to
rapidly isolate states in our two- and three-point correlation
functions.

\begin{table}[b]
	\centering
	\caption{The RMS radii for the various levels of smearing considered in this work.}
	\begin{ruledtabular}
	\begin{tabular}{cc}
		Sweeps of smearing & RMS radius (fm) \\
		\hline
		16  & 0.216 \\
		35  & 0.319 \\
		100 & 0.539 \\
		200 & 0.778 \\
	\end{tabular}
	\end{ruledtabular}
\label{tab:smearingRadii}
\end{table}

For the spin-flavour form of our local interpolator we choose to use,
\begin{align*}
	\chi_{\pi}(x) &= \overline{d}(x) \gamma_5 u(x) \\
	\chi_{\rho}(x) &= \overline{d}(x) \gamma_i u(x)
\end{align*}
for the $\pi$ and $\rho$ meson respectively.  When coupled with our
four smearing widths this allows for the construction of up to a 4
$\times$ 4 correlation matrix.  The use of the alternate bi-linear
forms $\gamma_0\gamma_5$ and $\gamma_0\gamma_i$ was considered, but
these were found to not provide any additional basis span when used
with more than two smearing levels.  We considered all combinations of
variational parameters $t_0$ and $\delta t$ in range 17-20 and 1-4
respectively where a superposition of states can be used to constrain
the analysis.  With regard to state isolation and the stability of the
analysis, the optimal choice is $t_0 = 17$ and $\delta t = 3$ for the
three heavier masses and $t_0 = 17$ and $\delta t = 2$ for the two
remaining lighter masses.  The use of an earlier $t_0$ value relative
to baryon studies \cite{Mahbub:2010,Menadue:2011,Mahbub:2012} on the
same ensembles is unsurprising given the larger energy gaps displayed
between the ground state and first excitation in the meson sector.

In performing our variational analysis we choose to use the
symmetrisation procedure outlined in Ref.~\cite{Mahbub:2013}.  Namely
we exploit the ensemble average symmetry
\[
	\cG_{ij}(\p, t) = \cG_{ji}(\p, t) \, ,
\]
and consider the improved unbiased estimator
\[
	\frac{1}{2} \left[ \cG_{ij}(\p, t) + \cG_{ji}(\p, t) \right]
\]
Enforcing this symmetry allows us to re-express the eigenvalue
equations \eqref{LeftEV} and \eqref{RightEV} in terms of a real
symmetric matrix
\[
	[G^{-1/2}(\p, t_0) G(\p, t_0 + \delta t) G^{-1/2}(\p,
          t_0)]_{ij} \, w^{\alpha}_j = \exp{-E_{\alpha}(\p) \delta t}
        \, w^{\alpha}_j \, . 
\]
The resulting eigenvectors from this formulation form an orthogonal
basis.  These orthogonal eigenvectors, $w^{\alpha}_i(\p)$, can also be
obtained from the eigenvectors $u^{\alpha}_i(\p)$ using
\[
	w^{\alpha}_i(\p) = G^{1/2}_{ij}(\p, t_0) \, u^{\alpha}_j(\p) \, .
\]
Upon normalising this basis, we are able to sort the eigenvectors
consistently across jackknife complement sets, and track eigenstates
across momenta using the approximate orthogonality condition
\[
	\vec{w}(\pp)^{\alpha} \cdot \vec{w}(\p)^{\beta} \simeq \delta^{\alpha\beta} \, .
\]
Correlators are normalized to be $\sim {\cal O}(1)$ by considering
\[
	\frac{1}{\sqrt{\cG_{ii}(\p, t_{src})}} \cG_{ij}(\p, t) \frac{1}{\sqrt{\cG_{jj}(\p, t_{src})}} \, .
\]
The resulting eigenvectors then give us a direct measure of the
relative contribution from each interpolator.  Accordingly, we 
also normalise the three-point correlators using the relevant 
two-point functions
\[
	\frac{1}{\sqrt{\cG_{ii}(\pp, t_{src})}} \cG^{\mu}_{ij}(\pp, \p, t_2, t_1) \frac{1}{\sqrt{\cG_{jj}(\p, t_{src})}} \, .
\]

\section{Results}

\subsection{Low lying meson spectrum}

In Figs.~1 and 2 we display the resulting spectrum below 3~GeV in the
$0^{-+}$ and $1^{--}$ channels respectively obtained from the $4
\times 4$ correlation matrix of smeared sources and sinks.  The masses
for the states considered in the form factors analysis are summarised
in Table~\ref{tab:mesonMasses}.

\begin{table*}[t]
	\centering
	\caption{Masses for the first two $\pi$ and $\rho$ meson
          eigenstates (ground state and first excitation) for various
          values of the hopping parameter, $\kappa$, extracted from
          our $4 \times 4$ correlation matrix of smeared sources and
          sinks.}
	\begin{ruledtabular}
	\begin{tabular}{llllll}
		$\kappa$ & $m_{\pi}^2$ (GeV${}^2$) & $m_{\pi}$ (GeV) & $m_{\rho}$ (GeV) & $m_{\pi^*}$ (GeV) & $m_{\rho^*}$ (GeV) \\
		\hline
		0.13700 & 0.3876(11) & 0.6226(9)   & 0.981(5)  & 1.66(7)  & 1.68(7)  \\
		0.13727 & 0.2647(9)  & 0.5145(9)   & 0.917(6)  & 1.59(4)  & 1.72(5)  \\
		0.13754 & 0.1509(7)  & 0.3884(9)   & 0.867(6)  & 1.55(8)  & 1.77(5)  \\
		0.13770 & 0.0811(6)  & 0.2848(11)  & 0.832(10) & 1.77(4)  & 1.81(7) \\
		0.13781 & 0.0260(10) & 0.1613(31)  & 0.793(14) & 1.50(23) & 1.77(11) \\
	\end{tabular}
	\end{ruledtabular}
\label{tab:mesonMasses}
\end{table*}

Beginning with the pion channel we find three well separated
eigenstates, consistent with the spectrum observed in previous works
examining the entire isovector meson sector
\cite{Dudek:2010,Gattringer:2008,Engel:2010}.  At the lightest mass we
find that our first excited state is consistent with the $\pi$(1300).
A notable feature in our spectrum is a significant shift in the mass
for the first excited state at the second lightest mass and a similar
jump for the second excited state at the middle mass.  This feature is
observed across the range of variational parameters considered and on
corresponding 3 $\times$ 3 analyses formed from subsets of the
variational basis.  A similar feature is observed in positive parity
spectrum of the nucleon \cite{Mahbub:2010,Mahbub:2013}. Examination of
the wave functions for these nucleon excitations \cite{Roberts:2013}
show significant finite volume effects for the lightest two masses
which may give rise to an increase in the eigenstate energies. It is
possible that we are observing a similar effect here.

\begin{figure}[t]
	\includegraphics[width=\columnwidth]{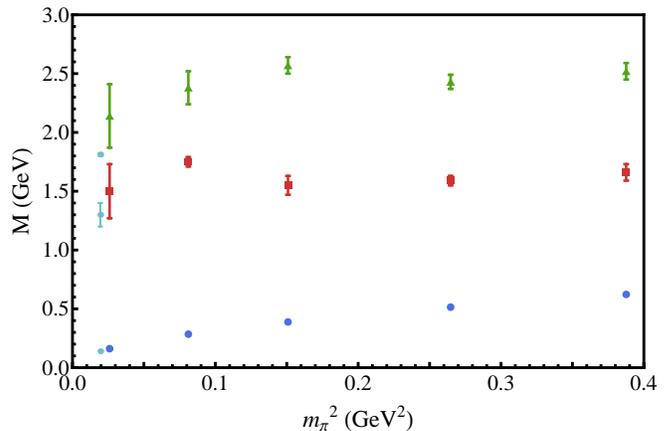}
	\caption{Low-lying eigenstates in the pion ($0^{+-}$) channel
          from our $4 \times 4$ correlation matrix of smeared sources
          and sinks.  The left-most points at the physical pion mass
          are experimental measurements of the spectrum.}
\end{figure}

In the rho meson channel, we again observe three well separated
eigenstates.  However, in this channel we expect to see two
eigenstates near 1600~MeV separated by about 250~MeV.  This suggests
that a basis of local operators is not able to isolate both of these
eigenstates.  A similar conclusion was found in
Ref.~\cite{Gattringer:2008}.  In Ref.~\cite{Engel:2010} they found
that the $\rho(1450)$ required a basis that contained displaced,
derivative operators.  Quark model expectations predict that for
these two states, one is $S$-wave dominant while the other $D$-wave
dominant \cite{Beringer:1900}.  Given the radial symmetry of our
operators, it is not possible to form $D$-wave states and so it
is not surprising that we are unable to isolate both states.  

The close agreement of our first excited state with the $\rho$(1700)
and the absence of the $\rho$(1450) from our spectrum suggests that
the $\rho(1700)$ is the $S$-wave state while the $\rho(1450)$ is the
$D$-wave state.  

Dudek {\it et al.} \cite{Dudek:2010} were able to successfully isolate
both of these states at larger quark masses and found a trend
consistent with our identification of states.  They found the $S$-wave dominated state to be
the lighter state at their heavier quark masses, and observed that
with decreasing quark mass the mass splitting between these two states
became smaller, with the states nearly degenerate at their lightest
mass of $m_{\pi} \simeq 400$~MeV.  A continuation of this trend to
lighter quark masses places $D$-wave dominated state lower in mass.

\begin{figure}[t]
	\centering
	\includegraphics[width=\columnwidth]{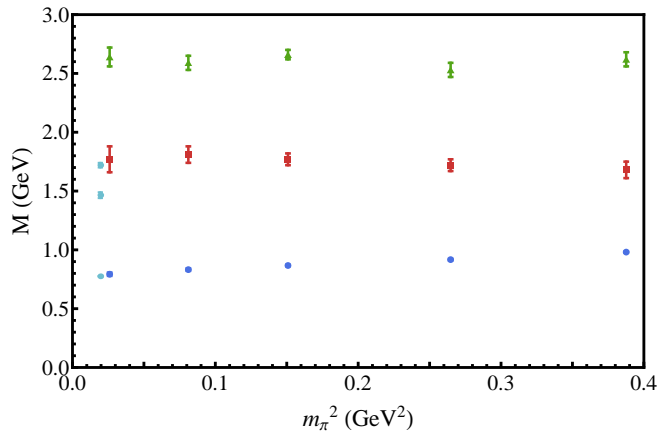}
	\caption{Low-lying eigenstates in the rho meson ($1^{--}$)
          channel from our $4 \times 4$ correlation matrix of smeared
          sources and sinks.  The left-most points at the physical
          pion mass are experimental measurements of the $\rho$-meson
          spectrum.}
\end{figure}

An important feature in the QCD spectrum is the possibility of
multi-particle intermediate states.  In the infinite-volume limit this
renders the majority of hadrons unstable under the strong interaction.
However on the finite-volume lattice, the QCD eigenstates are stable
and are composed of admixtures of both single-particle and
multi-particle states.

Some insight into the composition of states can be taken from the
physical spectrum and scattering thresholds.  However, the position of
these thresholds change on the finite volume.  Multi-particle states
are forced to overlap in the finite volume, giving rise to a
volume-dependent interaction energy.  Mixing with single-particle
dominated states further distorts the spectrum to the point where
intuition from infinite-volume scattering thresholds and the physical
spectrum becomes irrelevant, particularly in volumes with lattice
length $L \sim 3$ fm.

Below the finite-volume modified two-particle scattering threshold,
states are generally single-particle dominated but still contain
important contributions from nearby scattering channels.  The position
of states in the spectrum can be changed by varying the quark mass or
the volume of the lattice and the eigenstates can become maximally
mixed making their traditional identification as scattering states or
resonant states impossible.  In the case where a low-lying
finite-volume scattering threshold sits well below the resonant state,
then the lowest-lying state may be regarded as a two- or
multi-particle scattering state and the single-particle dominated
state is now the higher eigenstate.

In the case of the vector meson under study here, we must be careful
to ensure that the state we are exciting on the lattice is in fact the
single-particle dominated resonant state.  Though there is strong
evidence to suggest that local meson operators couple poorly to
scattering states, especially on larger volumes such as that under
investigation here, we perform a check to determine whether the
eigenstate isolated in our correlation-matrix analysis is single
particle in nature.

\begin{figure}[t]
	\centering
	\includegraphics[width=\columnwidth]{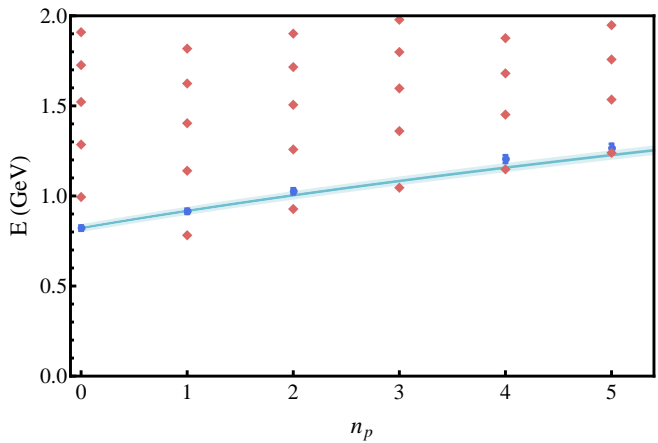}
	\caption{An example of our comparison between the dispersion
          relation $E = \sqrt{m^2 + n_p |\p_{min}|^2}$ (blue
          line), and the energies extracted from the
          finite-momentum correlators (blue circles) for the rho meson
          over a range of momenta.  Here $m$ is taken from the
          zero-momentum correlator, $|\p_{min}|$ is the magnitude of the
          lowest non-trivial momentum on the lattice.  The red diamonds
          provide the corresponding non-interacting $\pi\pi$-energies
          allowed by momentum conservation.  Results at the other
          values of $m_{\pi}$ are similar.}
\label{fig:dispersion}
\end{figure}

For all our ensembles, the ground state $\rho$ meson at rest is well below
the $\pi\pi$ threshold, and will be single-particle dominated.
However, upon applying the boost to momentum $\q$, the extracted
energy eigenstate now sits above the lowest-lying bare $\pi\pi$ energy
allowed by momentum conservation.  In order to determine whether the
state we have isolated in the boosted case is the finite-volume
$\rho$-meson or the lower-lying $\pi\pi$ scattering state, we compare
the extracted eigenstate energy against the expected energy given by
the dispersion relation for a single particle.

In Fig.~\ref{fig:dispersion} corresponding to the second-lightest
quark mass considered, we overlay the dispersion relation expectation
(blue band) with the energies extracted from the finite-momentum
correlators (blue data points) for a range of momenta.  Here we can
see that the energies extracted from the boosted correlator are in
excellent agreement with dispersion.  This is observed across all the
masses considered.

As a further check, we compare the non-interacting $\pi\pi$ energies
(red data points) allowed by momentum and parity conservation.  For
the single unit of lattice momentum relevant to our form factor
analysis, we find that the mass separation between the dispersion
result and the non-interacting $\pi\pi$ energy is significant.
Moreover, the attractive finite-volume interaction in the $\pi\pi$
system would act to further increase the separation between the single
and multi-particle dominated states.  As a matter of principle, we do
expect a $\pi\pi$ scattering state to reveal itself in the long
Euclidean-time tail of our correlation function.  However, our
interpolating fields have rendered this contribution to be negligible
at the finite-Euclidean times considered.

This indicates that our correlation functions are indeed dominated by
the resonant-like state of interest and not the lower-lying finite
volume scattering state.  Similar separations are found for all other
masses with the exception of the middle mass.  Through this process we
have determined that the state we have isolated in the boosted system
is in fact the state most closely related to the resonant $\rho$
meson.


\subsection{Form Factor Determination}

A key finding from our calculation of $g_A$ using the variational
approach \cite{Owen:2012}, was that the optimised operators obtained
from the correlation matrix analysis composed of various smeared
source/sink operators gave rise to significant improvement in the
quality and duration of plateaus from which the matrix elements were
extracted.  Comparison with a modest historically-typical choice of
smearing highlighted that excited state effects are significant and
suppress the value of $g_A$.

\begin{figure}
	\centering
	\subfigure[\ Charge form factor, $G_C$, for $m_{\pi}=296$~MeV.]{\includegraphics[width=\columnwidth]{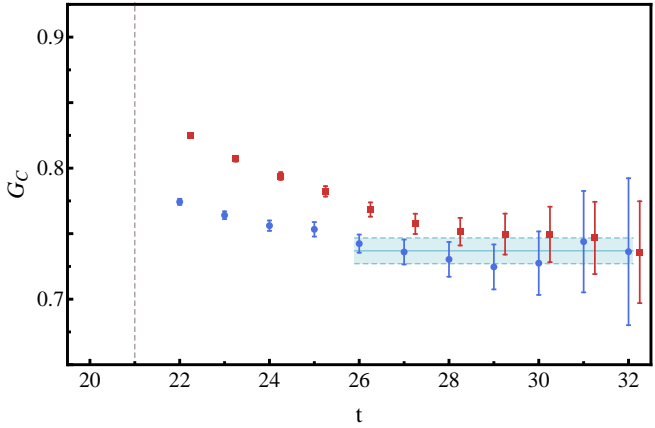}}
	\vspace{3mm}
	\subfigure[\ Magnetic form factor, $G_M$, for $m_{\pi}=156$~MeV.]{\includegraphics[width=\columnwidth]{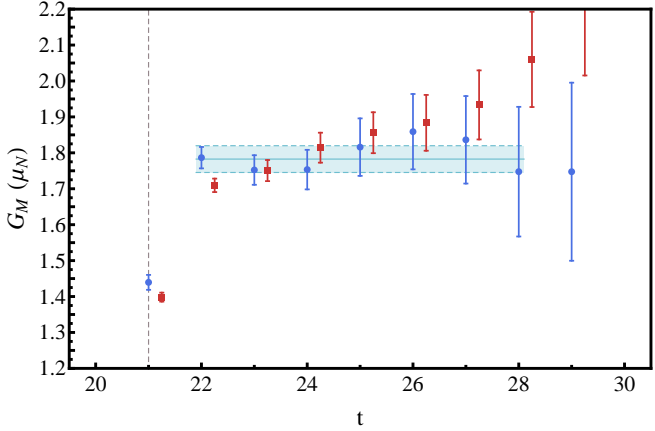}}
	\vspace{3mm}
	\subfigure[\ Quadrupole form factor, $G_Q$, for $m_{\pi}=296$~MeV.]{\includegraphics[width=\columnwidth]{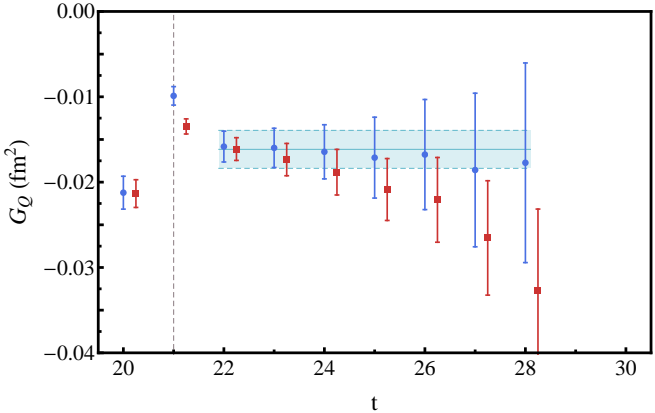}}
	\caption{A comparison of $\rho$-meson form factors as a
          function of Euclidean sink time for a single level of
          smearing and our variational approach.  The data sets
          obtained from ratios of three- and two-point functions are
          offset for clarity.  The (blue) circles denote the results
          from the variational approach while the (red) squares
          illustrate traditional results using the standard
          single-source method with a moderate level of smearing (35
          sweeps).  The vertical dashed line indicates the position of
          the current insertion.  The fitted value from the
          variational approach has been included (shaded band) to
          highlight where the single source approach is consistent
          with our improved method. }
\label{fig:plateaus}
\end{figure}

We are able to draw similar conclusions here.  In
Fig.~\ref{fig:plateaus} we present a comparison of the correlation
function ratios providing the rho-meson charge, magnetic and
quadrupole form factors using both the standard single-source approach
with a modest level of smearing and our variational approach.

The magnetic form factor, shown in Fig.~\ref{fig:plateaus}(b), is the
most striking example.  Here we see a clear difference in the quality
of the plateau.  For the correlation matrix approach, single-state
dominance follows the current at $t_c = 21$ immediately, allowing fits
as early as $t_S = 22$.  

For the standard approach, the excited states act to suppress the
value of $G_M$ at earlier time slices forcing one to wait until at
least $t_S = 24$ before an adequate $\chi^2_{dof}$ is obtained.
However, the central values show a systematic upward trend following this time 
slice and consequently there is no indication that a plateau has been obtained.

Similar conclusions can be drawn for the quadrupole form factor in
Fig.~\ref{fig:plateaus}(c).  Here we find that both the correlation
matrix and the standard approach give consistent values immediately
following the current, but diverge as we move out to later
time-slices.  Again there is a clear systematic drift in the results
obtained using the standard approach and it would be difficult to
select a fit region where the form factor can be determined with
confidence.  Given that we seek a region where the extracted form
factor is constant over successive time slices, we can clearly see the
improvement offered by the correlation matrix approach.

In the case of the charge form factor, Fig.~\ref{fig:plateaus}(a), the
two methods are in closer agreement and display a similar quality in
plateau.  In either case, some Euclidean time evolution is required
before a plateau is observed, but we find that the correlation matrix
approach gives a systematically lower value following the current and
plateaus a couple of time-slices earlier than the standard approach.

Though the examples presented are selected to highlight the
improvement using the variational method, we see significant
improvement across all masses and form factors considered.  Through
this method we consistently find that one is able to obtain
single-state dominance earlier and often with significant improvement
in the quality of the plateau.  Through the use of the variational
approach we are able to:
\begin{itemize}
\item Isolate an eigenstate at earlier Euclidean times,
\item Insert the conserved vector current at earlier Euclidean times,
\item Fit the correlation function at earlier Euclidean times,
\item Observe robust plateau behaviour,
\item Identify large Euclidean-time fit windows,
\item Determine the form factors with significantly reduced systematic
  errors, and
\item Determine the form factors with significantly reduced
  statistical uncertainties due to the admission of analysis at
  earlier Euclidean times.
\end{itemize}

\subsection{Form Factors and Static Quantities}

\begin{figure}[t]
	\centering
	\includegraphics[width=\columnwidth]{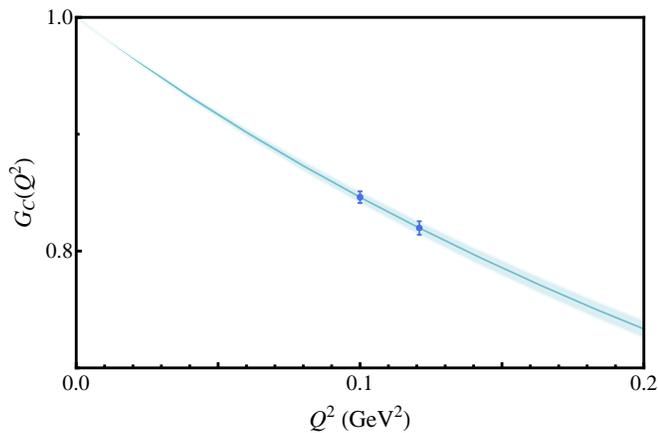}
	\caption{An example of the shift applied to the pion form
          factor to ensure that all eigenstates across all quark
          masses are at a common $Q^2$.  For the pion and its
          excitation, this shift is to $Q^2=0.10$~GeV${}^2$, while for
          the rho meson and its excitation we shift to
          $Q^2=0.16$~GeV${}^2$.}
\label{fig:FF_shift} 
\end{figure}

In extracting form factors across a wide range of quark masses, it is
important to note that for each mass there will be a slight change in
the value of $Q^2$ due to the variation in the mass of the hadron.  To
ensure that the comparison between quark masses and eigenstates is
meaningful, we make use of the monopole ansatz to shift the extracted
values for our form factors to a common $Q^2$.
Fig.~\ref{fig:FF_shift} demonstrates this shift for the pion.  For the
pion system, we shift $Q^2$ values to a common value of $Q^2 =
0.1$~GeV${}^2$, while for the $\rho$-meson system we select $Q^2 =
0.16$~GeV${}^2$.  These values are selected to minimise the shift for
the extractions at the lightest quark mass.  We use different values
to ensure that we minimise the shift for each system.  As the pion is
significantly lighter than the $\rho$ meson, a smaller value for $Q^2$
arises naturally.

Before we examine our form factor results, we note that all quantities
presented are the quark sector contributions for unit-charge quarks.
Here we choose to label these as the quark sector contributions to the
positive-charge eigenstate of the corresponding iso-triplet.  That
is, the quark contribution is labelled as the $u$-quark sector while
the anti-quark contribution is labelled as the $d$-quark sector.  As
we are working with exact isospin symmetry, these quark sector
contributions are equivalent and so we choose to present the $u$-quark
sector only.  As we are working in full QCD, the hadron form factor
will have contributions from the sea quarks.  However these
contributions have been neglected in our calculation.  Thus to make
meaningful comparison with experimental data, one should consider the
iso-vector quantities,
\[
	\langle 1_V \rangle = \frac{1}{2} ( \langle 1_+ \rangle - \langle 1_- \rangle ) \, ,
\]
where $\langle 1_{\pm} \rangle$ labels the positive and negative
members of the iso-triplet of the pion and $\rho$ meson.  However, under
exact isospin symmetry one finds that this iso-vector quantity defined
above is exactly that given by the connected quark sector contribution.

\subsubsection{Charge Form Factors}

\begin{figure}[t]
	\centering
	\includegraphics[width=\columnwidth]{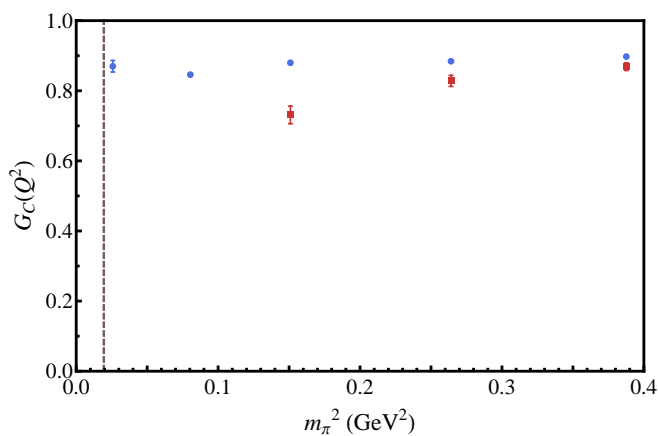}
	\caption{The unit-charge quark-sector contributions to the charge form factor, $G_C$, for the pion (blue
          circles) and its first excitation (red squares) at the
          common value $Q^2=0.10$~GeV${}^2$. The dashed line represents the physical point.} 
\label{fig:GC_pi_comb}
\end{figure}

\begin{figure}[t]
	\centering
	\includegraphics[width=\columnwidth]{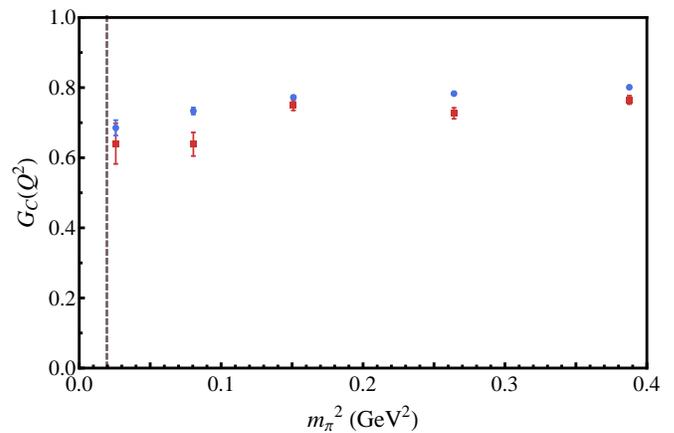}
	\caption{The unit-charge quark-sector contributions to the charge form factor, $G_C$, for the rho meson
          (blue circles) and its first excitation (red squares) at the
          common value $Q^2=0.16$~GeV${}^2$.} 
\label{fig:GC_rho_comb}
\end{figure}

In Figs.~\ref{fig:GC_pi_comb} and \ref{fig:GC_rho_comb} we display the
charge form factor $G_C$ for the $\pi$ and $\pi^*$ mesons and $\rho$
and $\rho^*$ mesons respectively.  In both channels we observe a
decrease in the charge form factor for the excitation which translates
to the excited states having a larger spatial extent.
Values are reported in Tables~\ref{tab:piGC} and \ref{tab:rhoGC}.

\begin{table}[t]
	\centering
	\caption{The quark sector contributions to the charge form
          factor, $G_C$, for the ground state and excited state pi
          meson, at the common $Q^2=0.10$~GeV${}^2$.  Results are for
          unit charge quarks.}
	\begin{ruledtabular}
	\begin{tabular}{lll}
	$m_{\pi}^2$ (GeV${}^2$) & $u_{\pi}$ & $u_{\pi^*}$ \\
	\hline
	0.3876(11)  & 0.898(1)  & 0.868(11)  \\
	0.2647(9)   & 0.884(2)  & 0.828(16)  \\
	0.1509(7)   & 0.880(3)  & 0.731(25)  \\
	0.0811(6)   & 0.846(5)  &     -      \\
	0.0260(10)  & 0.870(16) &     -      \\
	\end{tabular}
	\end{ruledtabular}
\label{tab:piGC}
\end{table}

\begin{table}[t]
	\centering
	\caption{The quark sector contribution to the charge form
          factor, $G_C$, of the ground state and excited state rho
          meson, at the common $Q^2=0.16$~GeV${}^2$.  Results are for
          unit charge quarks.}
	\begin{ruledtabular}
	\begin{tabular}{lll}
	$m_{\pi}^2$ (GeV${}^2$) & $u_{\rho}$ & $u_{\rho^*}$ \\
	\hline
	0.3876(11)  & 0.801(3)   & 0.765(12)    \\
	0.2647(9)   & 0.783(3)   & 0.727(16)    \\
	0.1509(7)   & 0.772(3)   & 0.749(14)    \\
	0.0811(6)   & 0.733(10)  & 0.639(34)    \\
	0.0260(10)  & 0.685(22)  & 0.640(58)    \\
	\end{tabular}
	\end{ruledtabular}
\label{tab:rhoGC}
\end{table}

\begin{figure}[t]
	\centering
	\includegraphics[width=\columnwidth]{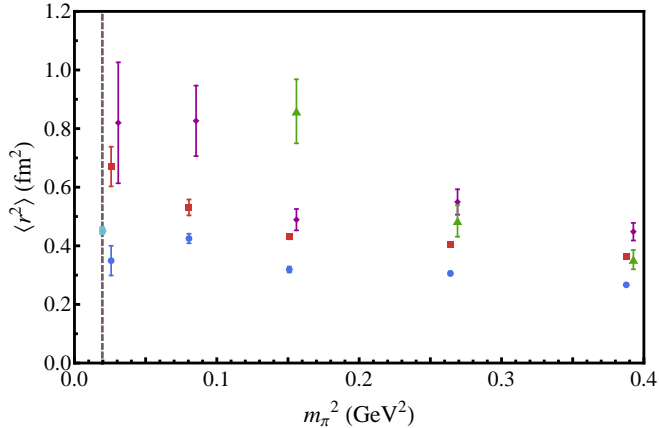}
	\caption{Mean-square charge radii, $\langle r^2 \rangle$, for
          the positive-charge states of the $\pi$ (blue circles),
          $\pi^*$ (green triangles), $\rho$ (red squares) and $\rho^*$
          (purple diamonds) mesons.  The grey dashed line represents
          the physical point, while the light blue data point is the
          experimental value for the pion from PDG
          \cite{Beringer:1900}. The $\pi^*$ and $\rho^*$ values have been 
          offset for clarity.}
\label{fig:charge_radii}
\end{figure}

To give us insight into the relative size of these states, we consider
the charge radii, shown in Fig.~\ref{fig:charge_radii}.  As was found
in \cite{Hedditch:2007}, the ground state vector meson is consistently
larger than the corresponding pseudoscalar meson.  It was noted that
this is consistent with quark model expectations where a hyperfine
interaction of the form $\frac{\vec{\sigma}_q \cdot
  \vec{\sigma}_{\bar{q}}}{m_q m_{\bar{q}}}$ is repulsive when spins
are aligned and attractive when spins are anti-aligned.  
In Fig.~10 we also include the experimental value for the pion radius
\cite{Beringer:1900}, which compares well with our determination.

For the heaviest quark masses where well determined values are
available for the excited states, it appears that a similar trend
holds between our $\rho^*$ and $\pi^*$ mesons.  The relatively large
radii for the excited states at light quark masses are interesting.
Results are tabulated in Table \ref{tab:radii}.

\begin{table}[t]
	\centering
	\caption{Mean-square charge radii ($\langle r^2 \rangle$) for
          the positive-charge states of the $\pi$ and $\rho$ mesons
          and their first excitations in units of fm${}^2$.}
	\begin{ruledtabular}
	\begin{tabular}{lllll}
	$m_{\pi}^2$ (GeV${}^2$) & $\pi$     & $\rho$    & $\pi^*$  & $\rho^*$ \\
	\hline
	0.3876(11)  & 0.267(4)  & 0.363(6)  & 0.35(3)  & 0.45(3)  \\
	0.2647(9)   & 0.306(7)  & 0.405(8)  & 0.48(5)  & 0.55(4)  \\
	0.1509(7)   & 0.320(10) & 0.430(8)  & 0.86(11) & 0.49(4)  \\
	0.0811(6)   & 0.425(16) & 0.531(27) &    -     & 0.83(12) \\
	0.0260(10)  & 0.349(51) & 0.670(68) &    -     & 0.82(21) \\
	\end{tabular}
	\end{ruledtabular}
\label{tab:radii}
\end{table}

\subsubsection{Magnetic Form Factors}

\begin{figure}[t]
	\centering
	\includegraphics[width=\columnwidth]{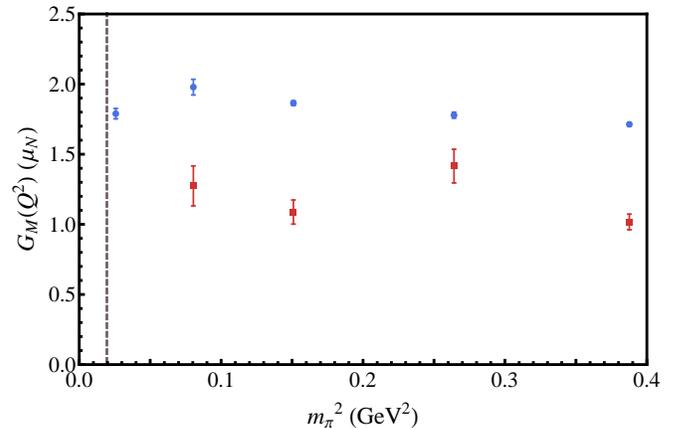}
	\caption{The unit-charge quark-sector contributions to the
          magnetic form factor, $G_M$, of the $\rho^+$ and $\rho^{*+}$
          mesons, at the common $Q^2=0.16$~GeV${}^2$.}
\label{fig:GM_rho_comb}
\end{figure}

The magnetic form factors for our $\rho$ and $\rho^*$ mesons are
illustrated in Fig.~\ref{fig:GM_rho_comb}.  For both the ground and
excited state we observe very little variation in the value as
$m_{\pi}$ varies.  As summarised in Table~\ref{tab:rhoGM}, we clearly
observe a significantly smaller value of $G_M$ for the excitation.
Though we would expect a decrease consistent with the decrease in the
charge form factor for this state, the degree of suppression suggests
that the magnetic moment for this state is smaller than the ground
state.  

\begin{figure}[t]
	\centering
	\includegraphics[width=\columnwidth]{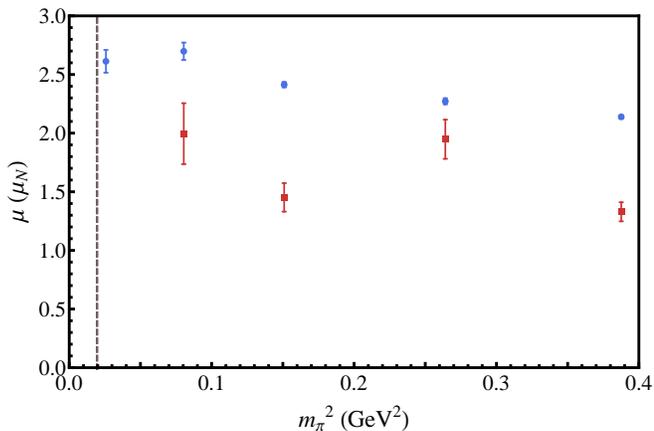}
	\caption{The quark-mass dependence of the $\rho^+$ and $\rho^{*+}$
          magnetic moments in units of the nuclear magneton.}
\label{fig:mu_rho_comb}
\end{figure}

In Table~\ref{tab:rhomu} we list magnetic moments for these
states where we invoke common scaling between the charge and magnetic
form factors.  We find a smaller moment for the $\rho^*$.
Our observation is consistent with results for meson magnetic moments
using the relativistic Hamiltonian \cite{Badalian:2012} approach.  In
particular, their value for $\frac{\mu_{2S}}{\mu_{1S}} \simeq 0.7$
compares well with our result of $0.74(9)$ for the second lightest mass. 
The quark-mass flow of these magnetic moments is illustrated in
Fig.~\ref{fig:mu_rho_comb}.

\begin{figure}[t]
	\centering
	\includegraphics[width=\columnwidth]{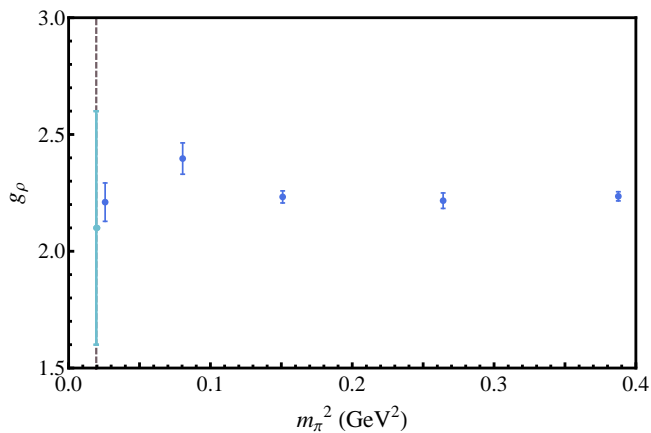}
	\caption{The $g$-factor of the $\rho$ meson is provided by the
          magnetic moment of the $\rho$ meson in natural magnetons.
          The dashed line highlights the physical point.  The light
          blue data point is the experimental determination of
          Ref.~\cite{Gudino:2013}.}
\label{fig:gfact}
\end{figure}

\begin{table}[t]
	\centering
	\caption{Unit-charge quark-sector contributions to the
          magnetic form factor of $\rho$ and $\rho^*$ mesons, at
          the common $Q^2=0.16$~GeV${}^2$.  The natural magneton has
          been converted to constant units of nuclear magnetons
          $\mu_{N}$.}
	\begin{ruledtabular}
	\begin{tabular}{lll}
	$m_{\pi}^2$ (GeV${}^2$) & $u_{\rho}\ (\mu_N)$ & $u_{\rho^*}\ (\mu_N)$ \\
	\hline
	0.3876(11)  & 1.713(12)  & 1.02(6)      \\
	0.2647(9)   & 1.779(21)  & 1.41(12)     \\
	0.1509(7)   & 1.864(17)  & 1.09(9)      \\
	0.0811(6)   & 1.978(56)  & 1.27(14)     \\
	0.0260(10)  & 1.791(37)  &    -         \\
	\end{tabular}
	\end{ruledtabular}
\label{tab:rhoGM}

	\centering
	\caption{Magnetic moments of the positive-charge states of the
          $\rho$ and $\rho^*$ mesons in units of nuclear magnetons
          $\mu_{N}$.}
	\begin{ruledtabular}
	\begin{tabular}{lll}
	$m_{\pi}^2$ (GeV${}^2$) & $\mu_{\rho}\ (\mu_N)$ & $\mu_{\rho^*}\ (\mu_N)$ \\
	\hline
	0.3876(11)  & 2.138(15)    & 1.33(8)        \\
	0.2647(9)   & 2.272(26)    & 1.94(17)       \\
	0.1509(7)   & 2.414(23)    & 1.45(12)       \\
	0.0811(6)   & 2.698(74)    & 2.00(26)       \\
	0.0260(10)  & 2.613(97)    &    -           \\
	\end{tabular}
	\end{ruledtabular}
\label{tab:rhomu}
\end{table}

In Fig.~\ref{fig:gfact} we show the $g$-factor for the $\rho$ meson,
provided by the magnetic moment of the $\rho$ meson in natural
magnetons.  Constituent quark model expectations suggest for a pure
$s$-wave state, that $g_{\rho} \simeq 2$.  Our result of $g_{\rho} =
2.21(8)$, taken from our lightest mass point, is larger than this
expectation and suggests a non-trivial value for the quadrupole moment
of the $\rho$ associated with $D$-wave mixing.  We observe a mild
downwards trend of the $g$-factor with increasing quark mass
suggesting that our results are compatible with the quark-model
expectation of $g_{\rho} = 2$ in the large quark mass limit.

Our results agree in value and behaviour with the previous quenched
determination \cite{Hedditch:2007} and dynamical study using
background field methods \cite{Lee:2008}.  In Fig.~\ref{fig:gfact} we
include the experimental determination of \cite{Gudino:2013}.  Within
the literature, the majority of model calculations
\cite{deMelo:1997,Bhagwat:2006,Aliev:2009,Roberts:2011,Pitschmann:2012,Djukanovic:2013,Mello:2014}
give a value of $g_{\rho}$ between $2.0$ and $2.4$ similar to our
determination of $2.21(8)$.

\subsubsection{Quadrupole Form Factors}

\begin{figure}[t]
	\centering
	\includegraphics[width=\columnwidth]{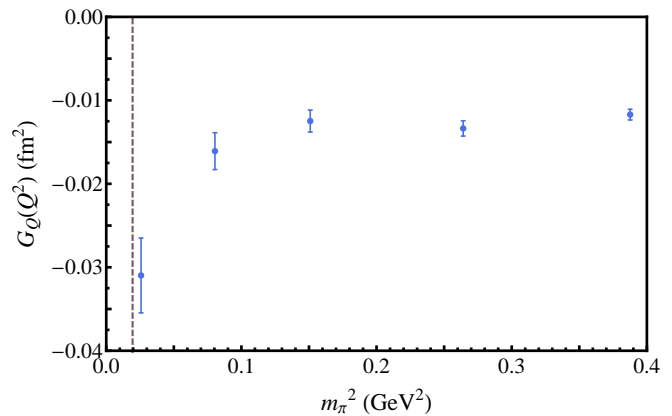}
	\caption{The unit-charge quark-sector contribution to the quadrupole form
          factor, $G_Q$, of the $\rho$ meson at the common
          $Q^2=0.16$~GeV${}^2$.  }
\label{fig:GQ_rho} 
\end{figure}

The quadrupole form factor for the $\rho$ meson is shown in
Fig.~\ref{fig:GQ_rho}.  For the excited $\rho^*$ meson, the signal was
too poor to extract a result.  As was found in the quenched study of
Ref.~\cite{Hedditch:2007} and the preliminary study of
Ref.~\cite{Gurtler:2008.PoS}, the quadrupole form factor is negative
in value.  The value of $G_{Q}$ varies mildly in the heavy quark regime,
however we observe a significant increase in the magnitude as we move
to the lightest mass.  Numerical results are provided in Table~\ref{tab:rhoGQ}.

\begin{figure}[t]
	\centering
	\includegraphics[width=\columnwidth]{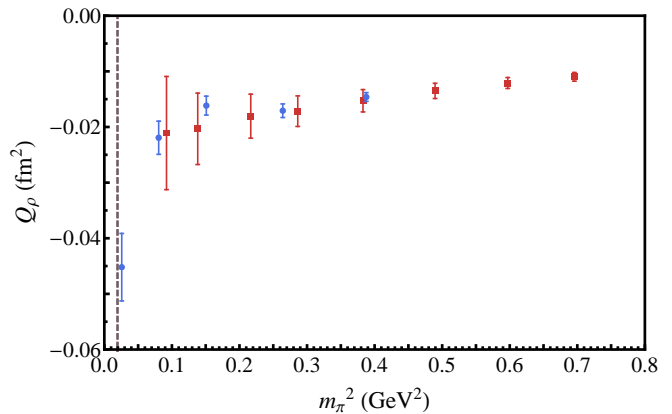}
	\caption{The quark-mass dependence of the $\rho$-meson
          quadrupole moment. The blue circles are from the current
          analysis, while the red squares are the quenched results
          from Ref.~\cite{Hedditch:2007}.  We see very similar
          behaviour between the results at the heavier masses.
          However, at our lightest mass we observe a significant
          increase in the magnitude of the quadrupole moment
          indicating a significant role of the pion cloud in the
          underlying structure of the rho meson.}
\label{fig:Q_rho}
\end{figure}

In Fig.~\ref{fig:Q_rho} we illustrate the quark mass dependence of the
quadrupole moment.  We also include the quadrupole moment
extracted \footnote{We note that an error was made in
  Ref.~\cite{Hedditch:2007} in converting the lattice form factors
  from natural units to fixed units.}  from the quenched data in
Ref.~\cite{Hedditch:2007}.  At heavier masses we find consistent
values and find a slight increase in the magnitude in the direction of
decreasing quark mass. However, we see a rapid increase in the
quadrupole moment magnitude at our lightest mass.  The value nearly
doubles in comparison with the next lightest mass. This indicates the
importance of light-quark dynamics to the underlying structure of the
rho meson giving rise to significant contributions from the pion
cloud. The dramatic variation observed warrants further investigation
into the chiral dynamics of this quantity, especially into the role
that finite volume effects may have on the determination.

\begin{table}[t]
	\centering
	\caption{Unit-charge quark-sector contributions to the
          quadrupole form factor of the $\rho$ meson, at the common
          $Q^2=0.16$~GeV${}^2$.}
	\begin{ruledtabular}
	\begin{tabular}{ll}
	$m_{\pi}^2$ (GeV${}^2$) &	$u_{\rho}\ (\mbox{fm}^2)$ \\
	\hline
	0.3876(11)  &	-0.0117(6)  \\
	0.2647(9)   &	-0.0134(9)  \\
	0.1509(7)   &	-0.0125(13) \\
	0.0811(6)   &	-0.0161(22) \\
	0.0260(10)  &	-0.0310(45) \\
	\end{tabular}
	\end{ruledtabular}
\label{tab:rhoGQ}
\end{table}

In Ref.~\cite{Lee:1965}, through considerations of the most general
free Lagrangian for a charged spin-1 system with minimal
electromagnetic coupling, it was shown that there exists an explicit
degree of freedom in the Lagrangian which can be parametrised by the
$g$-factor.  Consequently one finds that at tree-level the quadrupole
moment $Q_{\rho} = (1-g_{\rho})\nicefrac{e}{m_{\rho}^2}$.  With our value
of $g_{\rho} = 2.21(8)$, the tree-level value for $Q_{\rho} \simeq
-1.21(8) \nicefrac{e}{m_{\rho}^2}$.
In Table~\ref{tab:rhoQM} we report the quadrupole moment.  To make
contact with Ref.~\cite{Lee:1965}, we report results in terms of the
natural units of $e/m_\rho^2$ where $m_\rho$ is the mass of the $\rho$
meson observed on the lattice at each quark mass. At the lightest
quark mass considered we find $Q_{\rho} = -0.733(99)
\nicefrac{e}{m_{\rho}^2}$ indicating important contributions beyond tree
level, driven by the fundamental strong interactions of QCD.

\begin{table}[t]
	\centering
	\caption{The quadrupole moment of the $\rho^+$ meson in
          natural units of $e/m_{\rho}^2$, where $m_\rho$ is the mass
          of the $\rho$ meson observed on the lattice at each quark
          mass, and in fixed units of $\mbox{fm}^2$. }
	\begin{ruledtabular}	
	\begin{tabular}{lll}
	$m_{\pi}^2$ (GeV${}^2$) &	$Q_{\rho}^{\mathrm{nat}}$ & $Q_{\rho}\ (\mbox{fm}^2)$ \\
	\hline
	0.3876(11)  &	-0.362(20) & -0.0146(8)  \\
	0.2647(9)   &	-0.368(25) & -0.0171(12) \\
	0.1509(7)   &	-0.313(32) & -0.0162(17) \\
	0.0811(6)   &	-0.392(53) & -0.0219(30) \\
	0.0260(10)  &	-0.733(99) & -0.0452(61) \\
	\end{tabular}
	\end{ruledtabular}
\label{tab:rhoQM}
\end{table}

From the perspective of a non-relativistic quark model, the quadrupole
moment arises from an admixture of $S$ and $D$-wave components in the
wave function.  Thus our non-zero quadrupole moment, even at heavy masses,
indicates an important $D$-wave component to the $\rho$ meson. 

In a next-generation calculation where excited-state signals are
sufficiently accurate, the quadrupole moment can be used to determine
the dominant contributions to the $\rho(1450)$ and $\rho(1700)$ wave
functions. Thus, it offers the ideal tag to track these eigenstates
with varying quark mass.  In this way, one can determine if there is a
reordering of these states in the light quark regime.

\section{Summary and Conclusions}

We have established a general framework for the use of the variational
approach in the evaluation of hadronic form factors.  This approach
can be used to systematically eliminate excited state contributions to
ground state matrix elements.  It also allows one to evaluate the same
quantities for excited states with no additional effort.  

As we found in Ref.~\cite{Owen:2012}, the use of optimised
interpolators obtained from the variational approach results in rapid
isolation of the eigenstate, enabling earlier insertion of the probing
current.  Optimised interpolators at the sink result in rapid onset of
robust plateau behaviour enabling early and large Euclidean-time fit
windows.  Together these features act to reduce systematic errors
through the suppression of excited state contaminations and reduce
statistical uncertainties through the ability to insert the current
and establish fit windows earlier in Euclidean time.
This approach, coupled with the large lattice volume and light quark
masses, has resulted in an accurate determination of the pi and rho
electromagnetic form factors at the low $Q^2$.

Our light quark-mass determination of the rho-meson $g$-factor,
$g_{\rho} = 2.21(8)$, compares well with the experimental result of
\cite{Gudino:2013}, but with significantly smaller uncertainty.  This
value is consistent with earlier lattice and model evaluations, which
collectively prefer a $g$-factor slightly larger than the simple quark
model estimate of 2.  

As was found in the quenched calculation of Ref.~\cite{Hedditch:2007},
we obtain a negative value for the quadrupole form factor.  The onset
of significant chiral nonanalytic behaviour in the light quark-mass
regime is also observed.

Finally we have for the first time measured the electromagnetic form
factors for a light meson excitation.  
We find that the charge form factors for these states are smaller than
their ground state counterparts, consistent with expectations that
these states should be larger in size.  For the excited rho meson, we
observed a significantly smaller value for the magnetic form factor
and a smaller magnetic moment for the $\rho^*$.  Our observation of
$\mu_{\rho^*}/\mu_{\rho} = 0.74(9)$ supports the model prediction of
Ref.~\cite{Badalian:2012}.

Future work will investigate the finite-volume corrections to these
matrix element calculations drawing on the isolation of spin
polarizations in the lattice simulations and effective field theory in
relating the finite-volume matrix elements to those realised in
Nature.  

For the generalised eigenvalue problem with explicit spin degrees of
freedom on a finite-volume lattice, one can begin by quantifying the
volume-induced non-degeneracies in the spectrum.  For boosted
eigenstates with spin, there is interplay between the spin
polarization direction and the momentum direction.  For a fixed
momentum direction, the reduced rotational invariance induced by the
finite volume of the lattice gives rise to different multi-particle
dressings for different polarizations.  This gives rise to a subtle
variation in the energy of the eigenstates measured using different
polarizations.  To correctly account for this, a formalism to take the
infinite volume limit of each spin-momentum combination considered is
to be developed prior to combining the various components to extract
the form factors.  

Two-point function analyses are in the exploratory phase
\cite{OwenYoungEtal} and the effect is small.  Thus, an exploration of
the subtle nature of finite-volume corrections await a next-generation
simulation.  There the role of disconnected quark loop contributions
to the form factors can be ascertained.

\section*{Acknowledgements}

\sloppy We thank PACS-CS Collaboration for making their $2+1$ flavor
configurations available and acknowledge the important ongoing support
of the ILDG.  This research was undertaken with the assistance of
resources at the NCI National Facility in Canberra, Australia, and the
iVEC facilities at Murdoch University (iVEC@Murdoch) and the
University of Western Australia (iVEC@UWA).  These resources were
provided through the National Computational Merit Allocation Scheme,
supported by the Australian Government.  We also acknowledge eResearch
SA for their support of our supercomputers.  This research is
supported by the Australian Research Council.

\bibliography{vectorFF}







\end{document}